\documentclass[conference]{IEEEtran}
\IEEEoverridecommandlockouts
\usepackage[cmex10]{amsmath}

\usepackage{enumitem}
\usepackage{graphicx}
\usepackage{color}
\usepackage{amsmath}
\usepackage{mathtools}
\usepackage{multicol}
\usepackage{multirow}
\usepackage[english]{babel}
\usepackage{blindtext}
\usepackage{algorithm}
\usepackage{balance}
\usepackage{amsfonts}
\usepackage{bm}
\usepackage{stfloats}
\usepackage{subfig}
\usepackage{amsthm}
\usepackage{amssymb}
\usepackage{setspace}
\usepackage[nosort]{cite}
\usepackage{CJK}
\usepackage{cite}
\usepackage{caption}
\usepackage{comment}
\usepackage{array,multirow}
\usepackage{graphicx}

\theoremstyle{plain}

\usepackage[table]{xcolor}

\usepackage{algpseudocode}

\def\BibTeX{{\rm B\kern-.05em{\sc i\kern-.025em b}\kern-.08em
    T\kern-.1667em\lower.7ex\hbox{E}\kern-.125emX}}
\begin{document}
\captionsetup[figure]{labelformat={default},labelsep=period,name={Fig.}}

\title{A Bistatic Sensing System in Space-Air-Ground Integrated Networks\\
%\thanks{* Corresponding author.}
}

\author{
\IEEEauthorblockN{Xiangyu Li\IEEEauthorrefmark{1,2}, Bodong Shang\IEEEauthorrefmark{2}*, Qingqing Wu\IEEEauthorrefmark{1}}
    \IEEEauthorblockA{
    \IEEEauthorrefmark{1}Department of Electronic Engineering, Shanghai Jiao Tong University, Shanghai 200240, China\\
    \IEEEauthorrefmark{2}Eastern Institute for Advanced Study, Eastern Institute of Technology, Ningbo 315200, China\\
    e-mail: xiangyuli@sjtu.edu.cn, bdshang@eitech.edu.cn, qingqingwu@sjtu.edu.cn
    }
}

\maketitle

\begin{abstract}
Sensing is anticipated to have wider extensions in communication systems with the boom of non-terrestrial networks (NTNs) during the past years. In this paper, we study a bistatic sensing system by maximizing the signal-to-interference-plus-noise ration (SINR) from the target aircraft in the space-air-ground integrated network (SAGIN). We formulate a joint optimization problem for the transmit beamforming of low-earth orbit (LEO) satellite and the receive filtering of ground base station. To tackle this problem, we decompose the original problem into two sub-problems and use the alternating optimization to solve them iteratively. Using techniques of fractional programming and generalized Rayleigh quotient, the closed-form solution for each sub-problem is returned. 
Simulation results show that the proposed algorithm has good convergence performance.
% , and the increase of satellite transmit power and antenna number contributes positively to the target SINR performance.
Moreover, the optimization of receive filtering dominates the optimality, especially when the satellite altitude becomes higher, which provides valuable network design insights.
\end{abstract}


\section{Introduction}
\subsection{Motivations}
Sensing has been envisioned to become a key technology in the fifth generation (5G) and beyond wireless communication systems for obtaining emerging applications such as navigation and monitoring \cite{yazar20206g}. 
By applying sensing techniques, the quality of service (QoS) in terms of mutual communications as well as continuous detection and tracking of targets can be enhanced \cite{li2024sensing}.
% % wide sensing range is required
% Owing to the expansion of global networks and the boom of ubiquitous vehicles such as cars, aircraft, and even hot-air balloons, higher sensing quality is becoming a prerequisite for better traffic management, especially during emergent circumstances.
% Why need to sense aircraft
Due to the expansion of global networks and the advancement of air communication and transportation, vital significance has been brought to the task of sensing aircraft for aviation safety, traffic efficiency, and aerial security purposes \cite{shrestha2021survey}. 
By tracking and monitoring the movements of aircraft, not only the airspace traffic can be better controlled, but potential threats or unauthorized airspace entries can be detected promptly as well. Therefore, the enhancement of sensing quality for aircraft remains as a crucial problem.

% challenge in sensing due to propagation delay and path loss, thus design a bistatic sensing system which incorporate both terrestrial networks and non-terrestrial networks
% 战斗无人机
Traditionally, the implementation of target sensing is based on terrestrial networks (TNs) \cite{meng2023intelligent}. However, the direct extension of sensing to wider areas is facing challenges.
First, considering the large range and limited number of sensing equipment, it can be difficult to simultaneously monitor a large number of devices and timely focus on a certain target, especially in aerial regions. Second, long distances to the sensed targets will bring high propagation delay and path loss when echo signals are reflected back to the transceiving radar, as well as large investments in specific sensing equipment.

Satellite communications are recognized as a solution to satisfy the growing demand for seamless wireless connectivity and enhance the service availability in a global range \cite{10530195, shang2023coverage}. Compared with geostationary-earth orbit (GEO) and medium-earth orbit (MEO) satellites, low-earth orbit (LEO) satellites impose much less stringent requirements, including a lower latency, higher data rate and high visibility because of their lower altitudes \cite{liu2024max}. Thus, the LEO satellite can serve as a promising candidate for wide-range ubiquitous sensing, while satisfying the requirements of low-altitude economy.

% Why satellite, advantages
For sensing aircraft, LEO satellite-assisted sensing systems may have more advantages over terrestrial sensing systems. 
% wide coverage
On the one hand, an LEO satellite could simultaneously sense and detect target aircraft which are tens of kilometers from each other due to its wide coverage. The terrestrial radar base stations (BSs), however, mainly detect targets below their antenna deployment altitude, and those within their separate range of sector.
% energy consumption
On the other hand, powered by solar energy, satellites can have more energy for continuous sensing tasks, while the process of scanning and target searching for terrestrial radar BSs can be of huge energy consumption.

\subsection{Related Work}
% 侧重sensing应用场景的Related Work，因为本文的内容在于将老方法应用于新场景
Sensing techniques have been explored in many scenarios of wireless networks during the past decades, especially in TNs.
% V2X
The authors in \cite{cheng2022integrated} explore sensing in vehicle transportation settings and confirm the feasibility and necessity of using the functional integrated sensing and communication platform for different sensing modes.
% % Localization
% In \cite{peng2024semi}, the authors study a semi-passive intelligent reflecting surface (IRS)-enabled sensing system for localization of both the point target and the extended target under the influences of sensor number, reflecting element number and BS antenna number.
% Smart Manufacturing and IIoT
To monitor the operations of machines and the environment in the industrial Internet of Things (IIoT), Zhang \textit{et al.} use crowd sensing technology and propose a Particle Swarm Optimization-Elman Neural Network algorithm for position prediction \cite{zhang2023smart}.
% Remote sensing and environmental monitoring
Remote sensing enabled by satellite or airplane radar systems can be integrated with machine learning for the detection of airborne particulates and dust sources \cite{lary2016machine}, and be widely used in environmental monitoring \cite{messer2006environmental}.
% Human activity recognition or human computer interacting
Additionally, WiFi sensing is utilized indoor to recognize dynamic human activities such as breathing and heart beat by applying modeling or learning-based algorithms for over-the-air signal analysis \cite{ma2019wifi}.

% 加ziheng的ISAC文献，和tuanwei的sensing
Integrated sensing and communication (ISAC) systems have recently been explored for TNs. The work of \cite{zhang2023intelligent} studies intelligent omni surfaces (IOSs)-aided target sensing and user communication for a blocked urban region, where its bistatic setting for the radar system can effectively reduce the impacts of path loss.
Under the condition of imperfect channel state information (CSI), an intelligent reflecting surface (IRS)-aided sensing system is explored in the terrestrial setting \cite{tian2023active} while no receiving operations are employed. 
% 加两个ISAC in LEO Massive MIMO的文献
Furthermore, as satellite-assisted communications are becoming a new trend, \cite{you2022beam} investigates massive multiple-input multiple-output in ISAC LEO satellite systems using beam squint-aware techniques. The authors in \cite{huang2024secure} study non-orthogonal multiple access (NOMA)-aided precoding for ISAC by LEO satellites to ensure security.
It is worth mentioning that previous works on sensing systems are mainly targeted at TNs or non-terrestrial networks (NTNs) with direct space-ground channels.
There has not been a research on the detection of aerial targets in the space-air-ground integrated network (SAGIN).

\subsection{Contributions}
Motivated by the aforementioned works, we aim to use an LEO satellite as the transmitter and improve the sensing performance of a specific aircraft under the interfering sensing signals. Considering the huge path loss of echo signals reflected back from aircraft to the LEO satellite, we utilize a BS as the bistatic radar on the ground to receive the echo signals. 
The main contributions are summarized as follows.
\begin{itemize} 
\item We establish a bistatic sensing model for the SAGIN, which includes the satellite-to-aircraft channel, the aircraft-to-ground channel, and the sensing signal transmission model.
\item We jointly design the transmit beamforming and receive filtering to maxmize the SINR from the target aircraft under the total transmit power constraint and the receive filtering requirement.
\item We derive a closed-form solution for each sub-problem based on fractional programming and generalized Rayleigh quotient, respectively, to solve the proposed alternating problem efficiently.
\item The simulation results show that the receive filtering optimization dominates the overall performance, and higher transmit power and more antenna numbers contributes positively to the system improvement.
\end{itemize}

\textit{Notations:} Boldface, lower-case letters denote column vectors and boldface upper-case letters denote matrices. The superscript $(\cdot)^{-1}$, $(\cdot)^{T}$ and $(\cdot)^{H}$ represent the inverse, transpose and Hermitian, respectively. $\mathbf{I}_N$ refers to the identity matrices of $N$ dimensions, and $\mathbb{C}^{M\times N}$ denotes the $M\times N$ complex matrix. For a matrix
$\mathbf{A}$ and a vector $\mathbf{x}$, $\left[\mathbf{A} \right]_{i,j}$ stands for the $(i,j)$-th entry, while $\left[\mathbf{x} \right]_{i}$ represents the $i$-th entry.

\section{System Model}
\subsection{Channel Model}

\captionsetup{font={small}}
\begin{figure}[ht]
\begin{center}
\setlength{\abovecaptionskip}{+0.2cm}
\setlength{\belowcaptionskip}{-0.0cm}
\centering
  \includegraphics[width=2.8in, height=2.0in]{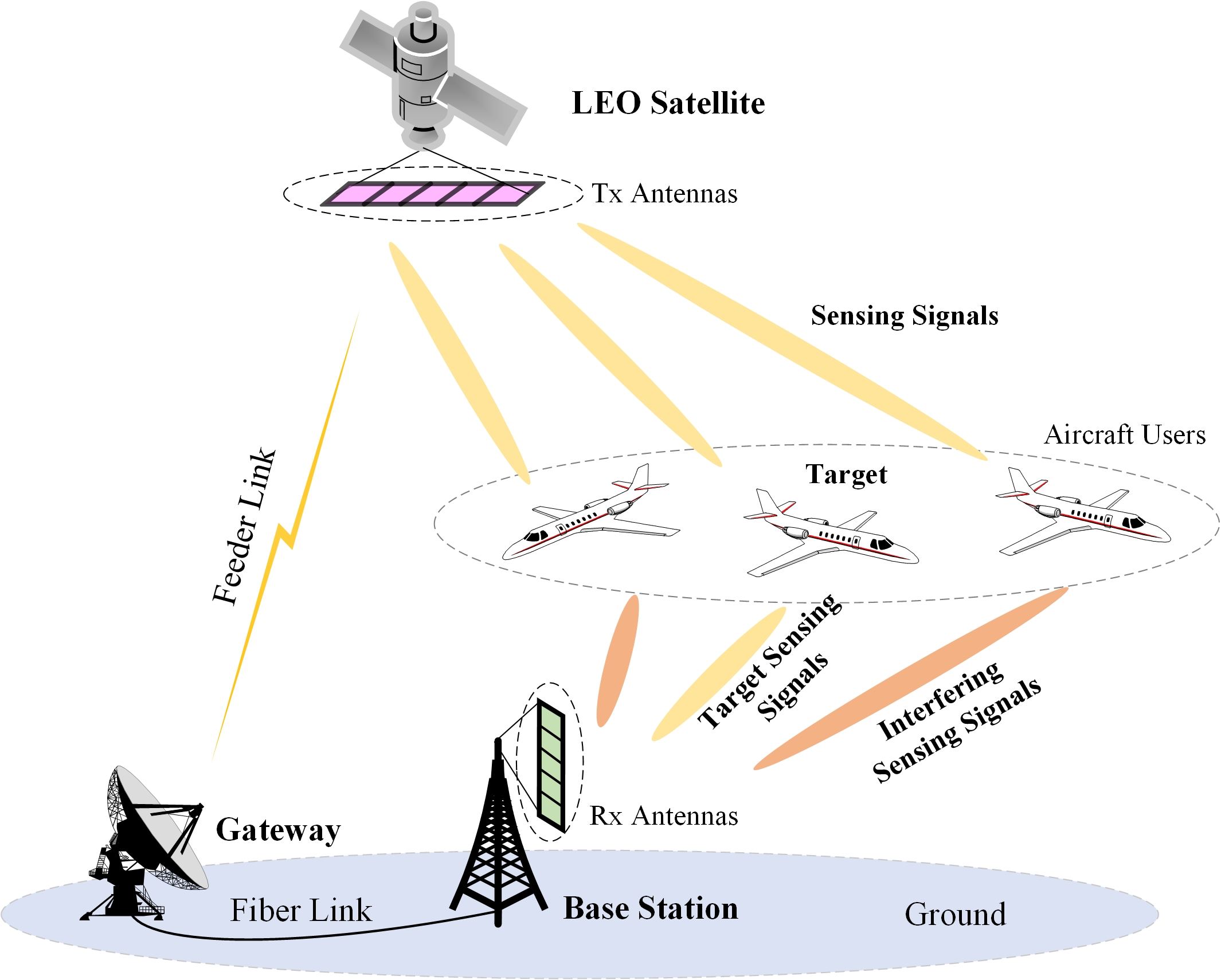}
\renewcommand\figurename{Fig.}
\caption{\small An illustration of the space-air-ground sensing system.}
\label{fig:System_Model}
\end{center}
\vspace{-7.5mm}
\end{figure}

As shown in Fig. \ref{fig:System_Model}, we consider a space-air-ground sensing system, consisting of a multi-antenna LEO satellite, a cluster of $K$ aircraft, a multi-antenna terrestrial BS, and a gateway.
Note that aircraft comprises but not limited to airplanes, unmanned aerial vehicles (UAVs), etc.
In order to mitigate the path-loss effect in the system, a bistatic radar is considered. The LEO satellite as a sensing signal transmitter is equipped with $M_t$ antennas and the BS as a sensing signal receiver is equipped with ${{M}_{r}}$ antennas. Assume that all aircraft are in the circular area of the same plane; the $k_0$-th aircraft is the sensed target while other $K-1$ aircraft are interference sources\footnote{Prior to optimization operations, the search for the existence of all aircraft in the designated area can be achieved by joint target search and communication channel estimation with omni-directional pilot signals \cite{liu2020joint}.}.
Moreover, the gateway is deployed as a control center for collecting CSI and performing optimization algorithms.

\subsubsection{Satellite-to-Aircraft (S2A) Channel} The channel between satellite and the $k$-th aircraft is denoted as ${\mathbf{h}_{S,k}}\in {{\mathbb{C}}^{1\times {{M}_{t}}}}$.
The $m_t$-th element of ${\mathbf{h}_{S,k}}$ is given as ${{\left[ {\mathbf{h}_{S,k}} \right]}_{{{m}_{t}}}}=\sqrt{\beta_{0}^{S2A}{{d}_{S,k}}^{-\alpha_{S2A}}} {{e}^{-j\left( {{m}_{t}}-1 \right)\frac{2\pi d}{\lambda }\sin \left( {\vartheta }_{S,k} \right)}}$, where $\beta_{0}^{S2A}$ is the path-loss at a reference distance for the S2A channel,
${{d}_{S,k}}$ stands for the distance between the satellite and the $k$-th aircraft, $\alpha_{SA}$ is the satellite-to-aircraft path-loss exponent, $d$ represents the antenna spacing, $\lambda$ indicates the wave length, and ${\vartheta}_{S,k}$ denotes the angle of arrival (AoA) from the transmitting antenna to the aircraft. Without loss of generality, we set the wave length as two times the antenna spacing, i.e. $\lambda =2d$.

\subsubsection{Aircraft-to-Ground (A2G) Channel} The uncorrelated Rician channel is used to model the channel responses, 
and the channel vector from the $k$-th aircraft to the BS is given by $\mathbf{h}_{k,{B}}=\sqrt{\beta_{k,B}}\mathbf{g}_{k,{B}}$, where ${\beta }_{k,B}=\beta_{0}^{A2G}{{d}_{k,B}}^{-\alpha_{A2G}}$ represents the large-scale fading coefficient, $\beta_{0}^{A2G}$ is the path-loss at a reference distance for the A2G channel, $\alpha_{AG}$ is the aircraft-to-ground path-loss exponent. The small-scale fading vector $\mathbf{g}_{k,B}\in {\mathbb{C}}^{{{M}_{r}}\times 1}$ is denoted as
\begin{equation}
    {\mathbf{g}_{k,B}}={\bar{\mathbf{g}}}_{k,B}\sqrt{\frac{\mathcal{K}}{\mathcal{K}+1}}+{\tilde{\mathbf{g}}}_{k,B}\sqrt{\frac{1}{\mathcal{K}+1}},
\end{equation}
where $\mathcal{K}$ denotes to the Rician factor, ${\bar{\mathbf{g}}}_{k,B}$ and ${\tilde{\mathbf{g}}}_{k,B}$ refer to the line-of-sight (LoS) and non-LoS (NLoS) components, respectively. Therefore, the A2G channel is written as
\begin{equation}
    {\mathbf{h}_{k,B}} = \sqrt{{{\beta}_{k,B}}}\left( {{\bar{\mathbf{g}}}}_{k,B}\sqrt{\frac{{\mathcal{K}}}{\mathcal{K}+1}}+{{\tilde{\mathbf{g}}}}_{k,B}\sqrt{\frac{1}{\mathcal{K}+1}} \right).
\end{equation}
We assume that 
${\left[ {{{\bar{\mathbf{g}}}}_{k,B}} \right]}_{{{m}_{r}}}={e}^{-j\left( {{m}_{r}}-1 \right)\frac{2\pi d}{\lambda }\sin \left( {{\vartheta }_{k,B}} \right)}$, where ${{\vartheta }_{k,B}}$ denotes the AoA from the aircraft to the terrestrial BS, while
the Rayleigh distributed term ${\tilde{\mathbf{g}}}_{k,B}$ describes the scattered multi-path and ${{\tilde{\mathbf{g}}}_{k,B}}\sim \mathcal{C}\mathcal{N}\left( {\mathbf{0}_{{{M}_{r}}}},{\mathbf{I}_{{M}_{r}}} \right)$.

\subsection{Sensing Signal Transmission}
Denote $\mathbf{t}\in {{\mathbb{C}}^{{{M}_{t}}\times 1}}$ as the transmit beamforming vector and use $q \in \mathcal{CN}(0,1)$ as the information symbol. The transmitted sensing signal $\mathbf{x}$ from the LEO satellite is given as $\mathbf{x}=\mathbf{t}q$, and the echo signal ${{y}_{k}}\in {{\mathbb{C}}^{1\times 1}}$ reflected from the $k$-th aircraft is ${{y}_{k}} = {\mathbf{h}_{S,k}} \mathbf{x} 
= {\mathbf{h}_{S,k}} \mathbf{t}q$. Then, the echo signal $\mathbf{y}_{e}\in {{\mathbb{C}}^{{{M}_{r}}\times 1}}$ received at the BS is expressed as
\begin{equation}
\begin{aligned}
\mathbf{y}_{e} 
% & =\sum\limits_{k=1}^{K}{{\mathbf{h}_{k,{{B}}}}{{y}_{k}}}+\bm{\omega}  \\ 
& ={\mathbf{h}_{{{k}_{0}},B}}{\mathbf{h}_{S,{{k}_{0}}}}\mathbf{t}q+\sum\limits_{k\ne k_0}^{K}{\mathbf{h}_{k,B}}{\mathbf{h}_{S,k}}\mathbf{t}q+\bm{\omega},
\end{aligned}
\end{equation}
Define the finite impulse response (FIR) filter at BS 
as $\mathbf{u}$, and $\mathbf{u}\in {{\mathbb{C}}^{{{M}_{r}}\times 1}}$.
Then, the filter output at the BS is given by
\begin{equation}
    y_o = \mathbf{u}^{H}\mathbf{y}_{e}.
\end{equation}
Finally, the SINR of the target aircraft is
\begin{equation}
    {\text{SINR}_{{{k}_{0}}}}=\frac{{\left| {\mathbf{u}^{H}}{\mathbf{h}_{{{k}_{0}},{B}}}{\mathbf{h}_{S,{{k}_{0}}}}\mathbf{t} \right|}^{2}}{{\mathbf{u}^{H}}\left( \sum\limits_{k\ne {{k}_{0}}}^{K}{{{\left| {\mathbf{h}_{k,{B}}}{\mathbf{h}_{S,k}}\mathbf{t} \right|}^{2}}}+{{\left| \bm{\omega} \right|}^{2}} \right)\mathbf{u}}.
    \label{Formula: SINR}
\end{equation}

\section{Problem Formulation and Solution}
In this paper, we assume the target $k_0$-th aircraft is in emergent conditions and higher QoS of sensing signal is required for better detection and localization.
Thus, we aim to maximize the $\text{SINR}_{k_0}$ of the echo signal from the target aircraft by jointly designing the transmit beamforming vector $\mathbf{t}$ and the receive filtering vector $\mathbf{u}$. 
The problem of interest can be written as
\begin{subequations}
\begin{equation}
    \begin{aligned}
        \mathcal{P}1: 
        \mathop{\mathrm{max}}\limits_{\boldsymbol{\mathrm{t},\mathrm{u}}} \quad f_1\left(\mathbf{t},\mathbf{u}\right)=\text{SINR}_{k_0}
    \end{aligned}
\end{equation}
\begin{equation}
    \text{s.t.} \quad \mathrm{R}1: \mathbf{t}^{H} \mathbf{t} \leq P_t,
    \label{R1}
\end{equation}
\begin{equation}
    \quad \quad \mathrm{R}2: \mathbf{u}^{H} \mathbf{u} = 1,
    \label{R2}
\end{equation}
\end{subequations}
where $\mathrm{R}1$ is the sensing power budget requirement, $\mathrm{R}2$ is the receive filtering requirement. Note that the problem ($\mathcal{P}1$) is a non-convex fractional programming expression which requires further operation to get the final optimality. In the following, we will divide the problem ($\mathcal{P}1$) into two sub-problems to solve it based on the alternating optimization.

\subsection{Transmit Beamforming Vector Design}
For fixed $\mathbf{u}$, we first introduce an auxiliary variable $\varpi$ to track the closed-form solution of $\mathbf{t}$. According to \cite{guo2020weighted,dampahalage2021weighted}, the fractional programming problem ${\max }_{x}\left\{f\left( x \right)/g\left( x \right)\right\}$ can be converted into ${\max }_{\left\{ x,y \right\}}\left\{ 2y\sqrt{f\left( x \right)}-{{y}^{2}}g\left( x \right) \right\}$, where $y$ is a new auxiliary variable. 

Denote $\mathbf{q}=\mathbf{u}^{H}{\mathbf{h}_{{{k}_{0}},{B}}}{\mathbf{h}_{S,{{k}_{0}}}}$ and $\mathbf{r}=\mathbf{u}^{H}{\mathbf{h}_{k,{B}}}{\mathbf{h}_{S,k}}$, where $\mathbf{q}\in {{\mathbb{C}}^{1\times {M}_{t}}}$ and $\mathbf{r}\in {{\mathbb{C}}^{1\times {M}_{t}}}$. With an auxiliary variable $\varpi$, the problem ($\mathcal{P}1$) can be rewritten as
\begin{subequations}
\begin{equation}
    \begin{aligned}
        \mathcal{P}2:       \mathop{\mathrm{max}}\limits_{\mathbf{t},\varpi,\mathbf{u}} f_2\left(\mathbf{t},\varpi;\mathbf{u}\right)
    \end{aligned}
\end{equation}
\begin{equation}
    \text{s.t.} (\mathrm{\ref{R1}}), (\mathrm{\ref{R2}}),
\end{equation}
\end{subequations}
where the new objective function is
\begin{equation}
\begin{aligned}
& f_2\left(\mathbf{t},\varpi;\mathbf{u}\right) \\
& = 2\mathcal{R}\left\{\varpi^H \mathbf{q}\mathbf{t} \right\} -\left|\varpi\right|^{2} \left( \sum\limits_{k\ne k_0}^{K}{{{\left| \mathbf{r}\mathbf{t} \right|}^{2}}}+{{\left| \mathbf{u}^{H}\bm{\omega} \right|}^{2}} \right), 
\end{aligned}
\label{Formula: P2}
\end{equation}
and $\varpi$ is updated according to \cite{shen2018fractional}, which is expressed as
\begin{equation}
    \varpi =\frac{\mathbf{q}\mathbf{t}}{\sum\limits_{k\ne {{k}_{0}}}^{K}{{{\left| \mathbf{r}\mathbf{t} \right|}^{2}}}+{{\left| {\mathbf{u}^{H}} \bm{\omega} \right|}^{2}}}.
    \label{Formula: varpi}
\end{equation}
Accordingly, the optimal $\mathbf{t}^{*}$ can be expressed as
\begin{equation}
    {\mathbf{t}^{*}}=\underset{\mathbf{t},\varpi;\mathbf{u}}{\mathop{\arg \max }}\,{{f}_{2}}\left( \mathbf{t},\varpi;\mathbf{u} \right).
\end{equation}
The Lagrangian of the associated problem ($\mathcal{P}2$) is defined as
\begin{equation}
    \mathcal{L}\left(\mathbf{t},\varpi,\lambda;\mathbf{u} \right)=f_2\left( \mathbf{t},\varpi;\mathbf{u} \right)-\lambda \left( {\mathbf{t}^{H}}\mathbf{t}-1 \right).
\end{equation}
Based on the Lagrangian, four conditions, also known as Karush-Kuhn-Tucker (KKT) conditions, are listed as
\begin{subequations}
\begin{equation}
    {\mathbf{t}^{H}}\mathbf{t}-1=0,
\end{equation}
\begin{equation}
    \lambda \ge 0,
\end{equation}
\begin{equation}
    \lambda \left( {\mathbf{t}^{H}}\mathbf{t}-1 \right)=0,
\end{equation}
\begin{equation}
    {{\nabla }_{t}}\mathcal{L}\left( \mathbf{t},\varpi ,\lambda ;\mathbf{u} \right)=2\varpi {\mathbf{q}^{H}}-2{\bm{\Xi} }\mathbf{t}-2\lambda {\mathbf{I}_{{{M}_{t}}}}\mathbf{t}=0,
\end{equation}
\end{subequations}
where $\bm{\Xi} ={{\varpi }^{2}}\sum\limits_{k\ne {{k}_{0}}}^{K}{\mathbf{r}^{H}\mathbf{r}}\in {{\mathbb{C}}^{{{M}_{t}}\times {{M}_{t}}}}$. Then, the closed-form solution can be expressed as 
% ${\mathbf{t}^{*}}=\varpi {{\left( \bm{\Xi} +\lambda {\mathbf{I}_{{{M}_{t}}}} \right)}^{-1}}{\mathbf{q}^{H}}$. 
\begin{equation}
    {\mathbf{t}^{*}}=\varpi {{\left( \bm{\Xi} +\lambda {\mathbf{I}_{{{M}_{t}}}} \right)}^{-1}}{\mathbf{q}^{H}}.
    \label{Formula: t_solution}
\end{equation}

The optimal dual variable $\lambda$ can be searched efficiently via bisection with appropriate upper and lower bound \cite{boyd2004convex}. The calculation of $\mathbf{t}$ is represented in Algorithm 1 based on the bisection method, where $\varepsilon_1$ and $\varepsilon_2$ are tolerance values.

\begin{algorithm}[t]
	\caption{Transmit Beamforming Vector Update Algorithm} %算法的名字
	\begin{algorithmic}[1]
		\State$\textbf{Input}$: $\lambda_{\text{min}}$, $\lambda_{\text{max}}$, and $\lambda = \frac{1}{2}\left(\lambda_{\text{min}}+\lambda_{\text{max}}\right)$
		\Repeat
        \State update $\mathbf{t}$ using equation (\ref{Formula: t_solution})
        \If {$\|\mathbf{t}\|> P_d$}
        \State $\lambda_\text{min}=\lambda$
        \Else
        \State $\lambda_\text{max}=\lambda$
        \EndIf
        \Until $\left| \lambda_{\text{min}}-\lambda_{\text{max}} \right| < \varepsilon_1$ and $\left| \|\mathbf{t}\| - P_d \right| < \varepsilon_2 $.
	\end{algorithmic}
\end{algorithm}

\subsection{Receive Filtering Vector Design}
When $\mathbf{t}$ is fixed, we design the solution of $\mathbf{u}$ for ($\mathcal{P}1$) using the generalized Rayleigh quotient method and the stagnation point method \cite{zhang2023intelligent}.

Denote $\mathbf{N}={\left| {\mathbf{h}_{{{k}_{0}},{B}}}{\mathbf{h}_{S,{{k}_{0}}}}\mathbf{t} \right|}^{2}$ and $\mathbf{D}=\sum\limits_{k\ne k_0}^{K}{{{\left| {\mathbf{h}_{k,{B}}}{\mathbf{h}_{S,k}} \mathbf{t} \right|}^{2}}}+{{\left| \bm{\omega} \right|}^{2}}$, the optimal $\mathbf{u}^{*}$ can be written as
\begin{equation}
    {\mathbf{u}^{*}}=\underset{\mathbf{u}}{\mathop{\arg \max }}\,\frac{\mathbf{u}^{H}\mathbf{N}\mathbf{u}}{\mathbf{u}^{H}\mathbf{D}\mathbf{u}},
    \label{Formula: u_opt_1}
\end{equation}
where $\mathbf{N}$ is a Hermitian matrix while $\mathbf{D}$ is a positive definite Hermitian matrix. 
Consider a unitary matrix $\mathbf{V}$ which satisfies $\mathbf{V}^H=\mathbf{V}^{-1}$ and $\mathbf{V}^H\mathbf{V}=\mathbf{I}_{M_r}$, we can re-express $\mathbf{D}$ as $\mathbf{D}=\mathbf{V}\text{diag}\left( {{\lambda }_{1}},{{\lambda }_{2}},\ldots,{\lambda }_{{M}_{r}} \right){\mathbf{V}^{H}}$, and denote $\mathbf{L}=\mathbf{V}\text{diag}\left( \sqrt{{\lambda }_{1}},\sqrt{{\lambda }_{2}},\ldots,\sqrt{{\lambda }_{{M}_{r}}} \right){\mathbf{V}^{H}}$, where for $i=1,2,\ldots,M_r$, ${\lambda }_{i}$ are positive eigenvalues of $\mathbf{V}$. Based on eigenvalue decomposition theory, we have $\mathbf{D}=\mathbf{L}\cdot\mathbf{L}$. Take $\mathbf{p}=\mathbf{L}\mathbf{u}$, the original objection function can be expressed as an equivalent form of
\begin{subequations}
\begin{equation}
    \mathcal{P}3: \quad  \mathop{\mathrm{max}}\limits_{\boldsymbol{\mathrm{p}}} \quad {\mathbf{p}^{H}}{\mathbf{L}^{-1}}\mathbf{N}{\mathbf{L}^{-1}}\mathbf{p}
\end{equation}
\begin{equation}
    \text{s.t.} \quad \mathbf{p}^{H} \mathbf{p} = C,
\end{equation}
\end{subequations}
where $C$ is a constant which is independent of the final optimality.
To attain its optimal solution, the Lagrangian $L\left( \mathbf{p},\nu  \right)={\mathbf{p}^{H}}{\mathbf{L}^{-1}}\mathbf{N}{\mathbf{L}^{-1}}\mathbf{p}-\nu \left( {\mathbf{p}^{H}}\mathbf{p}-C \right)$ is introduced. By following the optimality condition ${{\nabla }_\mathbf{p}}\mathcal{L}\left( \mathbf{p},\nu  \right)=2{\mathbf{L}^{-1}}\mathbf{N}{\mathbf{L}^{-1}}\mathbf{p}-2\nu \mathbf{p}=0\Rightarrow {\mathbf{L}^{-1}}\mathbf{N}{\mathbf{L}^{-1}}\mathbf{p}=\nu \mathbf{p}$, the optimal solution $\mathbf{p}^{*}$ is the eigenvector corresponding to the maximum eigenvalue of ${\mathbf{L}^{-1}}\mathbf{N}{\mathbf{L}^{-1}}$. Then, the closed-form analytical solution of the previous equation (\ref{Formula: u_opt_1}) is 
% $\mathbf{u}^{*}=\mathbf{L}^{-1}\mathbf{p}^{*}$ 
\begin{equation}
    % \mathbf{u}^{*}=\mathbf{L}^{-1}\mathbf{p}^{*}\sqrt{\frac{C}{\|\mathbf{p}^{*}\|_{2}^{2}}},
    \mathbf{u}^{*}=\frac{\mathbf{L}^{-1}\mathbf{p}^{*}}{\|\mathbf{p}^{*}\|},    
    \label{Formula: u_solution}
\end{equation}
with the attained $\mathbf{p}^{*}$. Finally, the maximum SINR of the sensing signal from the target $k_0$-th aircraft is represented as
\begin{equation}
\text{SINR}_{{{k}_{0}}}^{*}=\frac{{{\left( {\mathbf{u}^{*}} \right)}^{H}}\mathbf{N}{\mathbf{u}^{*}}}{{{\left( {\mathbf{u}^{*}} \right)}^{H}}\mathbf{D}{\mathbf{u}^{*}}}.
\end{equation}

\subsection{Algorithm Summary and Complexity Analysis}
Combining the above derivations, the proposed alternating optimization from joint transmit beamforming and receive filtering is summarized in Algorithm 2, where $\varepsilon_3$ is the tolerance value for iterated SINR.

Thereinto, with the matrix inverse operation and the denoted iteration number $I_{\lambda}$ of the bisection search for $\lambda$, the computational complexity to update $\mathbf{t}$ is approximately calculated as $\mathcal{O}\left(I_{\lambda}{M_r}\right)$; while the update of $\mathbf{u}$, which comprises an eigenvalue decomposition, requires the complexity of $\mathcal{O}\left({M_r}^3\right)$. Hence, the complexity of the proposed algorithm at each iteration is $\mathcal{O}\left(I_{\lambda}{M_r}+{M_r}^3 \right)$.

\begin{algorithm}[t]
	\caption{Alternating Optimization Algorithm for $\mathcal{P}1$} %算法的名字
	\begin{algorithmic}[1]
		\State$\textbf{Initialize}$: $\mathbf{t}$, $\mathbf{u}$, set $i=0$, $\text{SINR}^{(0)}=0$
		\Repeat
        \State update $\varpi$ using equation (\ref{Formula: varpi})
        \State update $\mathbf{t}$ using equation (\ref{Formula: t_solution})
        \State update $\mathbf{u}$ using equation (\ref{Formula: u_solution})
        \State $i = i + 1$
        \State calculate $\text{SINR}^{(i)}$ with equation (\ref{Formula: SINR})
        \Until $\left| \text{SINR}^{(i)} - \text{SINR}^{(i-1)} \right| < \varepsilon_3 $.
	\end{algorithmic}
\end{algorithm}

% \captionsetup{font={small}}
% \begin{figure}[ht]
% \begin{center}
% \setlength{\abovecaptionskip}{+0.0cm}
% \setlength{\belowcaptionskip}{-0.4cm}
% \centering
%   \includegraphics[width=3.0in, height=2.2in]{figures/Fig1_conv_mk.eps}
% \renewcommand\figurename{Fig.}
% \caption{\small Convergence validation.}
% \label{fig1}
% \end{center}
% \end{figure}

\section{Simulation Results}
% \subsection{Simulation Scenario}
In this section, simulation results are provided to validate the proposed algorithm in the studied bistatic space-air-ground sensing system. Unless otherwise noted, we set the S2A channel path-loss $\alpha_{S2A}=2.0$, the A2G channel path-loss $\alpha_{A2G}=2.2$, and the noise power is assumed to be $1\times10^{-12}$ W. Moreover, the LEO satellite and the BS are set to be located at $\left( -10,20,300 \right)$ km and $\left( 0,0,0 \right)$ km, respectively. A total of $K=4$ aircraft, including one target aircraft and three other interfering aircraft, are uniformly distributed in a circular aerial region with a radius of $10$ km centering at $\left(20,20,10\right)$ km, i.e. all of them are at an altitude of $10$ km above the ground.
In addition, four benchmarks are provided to evaluate the performance of the proposed algorithm: 1) Optimization of receive filtering vector alone; 2) Optimization of the transmit beamforming vector alone; 3) Random receive filtering vectors and transmit beamforming vectors; 4) Transceiving mode where LEO satellite transmits and receives the sensing signal.

% \captionsetup{font={small}}
% \begin{figure}[ht]
% \begin{center}
% \setlength{\abovecaptionskip}{+0.0cm}
% \setlength{\belowcaptionskip}{-0.4cm}
% \centering
%   \includegraphics[width=3.0in, height=2.2in]{figures/Fig1_conv_mk.eps}
% \renewcommand\figurename{Fig.}
% \caption{\small Convergence validation.}
% \label{fig1}
% \end{center}
% \end{figure}

% \captionsetup{font={small}}
% \begin{figure}[ht]
% \begin{center}
% \setlength{\abovecaptionskip}{+0.0cm}
% \setlength{\belowcaptionskip}{-0.4cm}
% \centering
%   \includegraphics[width=3.0in, height=2.2in]{figures/Fig2_Altitude.eps}
% \renewcommand\figurename{Fig.}
% \caption{\small Optimal SINR versus satellite altitude.}
% \label{fig2}
% \end{center}
% \end{figure}

% \captionsetup{font={small}}
% \begin{figure}[ht]
% \begin{center}
% \setlength{\abovecaptionskip}{+0.0cm}
% \setlength{\belowcaptionskip}{-0.4cm}
% \centering
%   \includegraphics[width=3.0in, height=2.2in]{figures/Fig3_Antenna_Number.eps}
% \renewcommand\figurename{Fig.}
% \caption{\small Optimal SINR versus satellite antenna number.}
% \label{fig3}
% \end{center}
% \end{figure}

% ***********************************************

\begin{figure*}
\noindent
\begin{minipage}{0.33\textwidth}
  \centering
  \includegraphics[width=2.55in, height=1.75in]{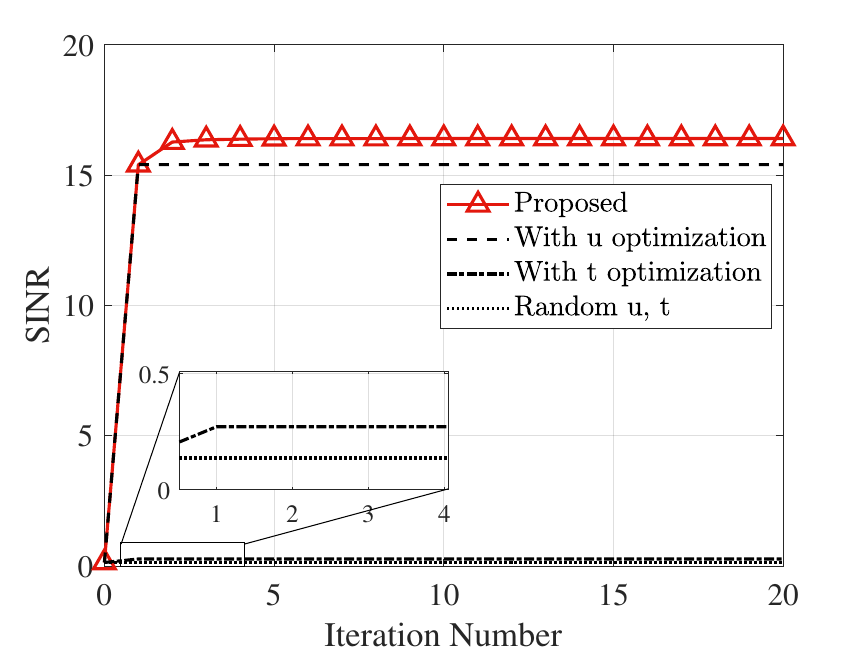}
  \vspace{-3mm}
  \renewcommand\figurename{Fig.}
  \caption{\footnotesize Convergence validation.}
  \label{fig1}
\end{minipage}
\begin{minipage}{0.33\textwidth}
  \centering
  \includegraphics[width=2.55in, height=1.75in]{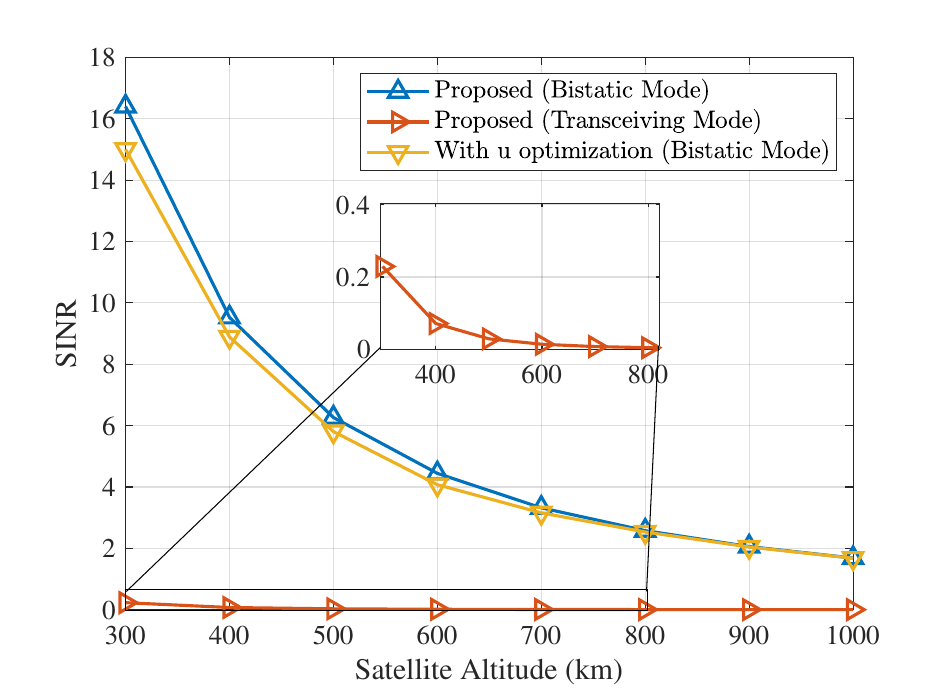}
  \vspace{-3mm}
  \renewcommand\figurename{Fig.}
  \caption{\footnotesize Optimal SINR \textit{VS} satellite altitude.}
  \label{fig2}
\end{minipage}
\begin{minipage}{0.33\textwidth}
  \centering
  \includegraphics[width=2.55in, height=1.75in]{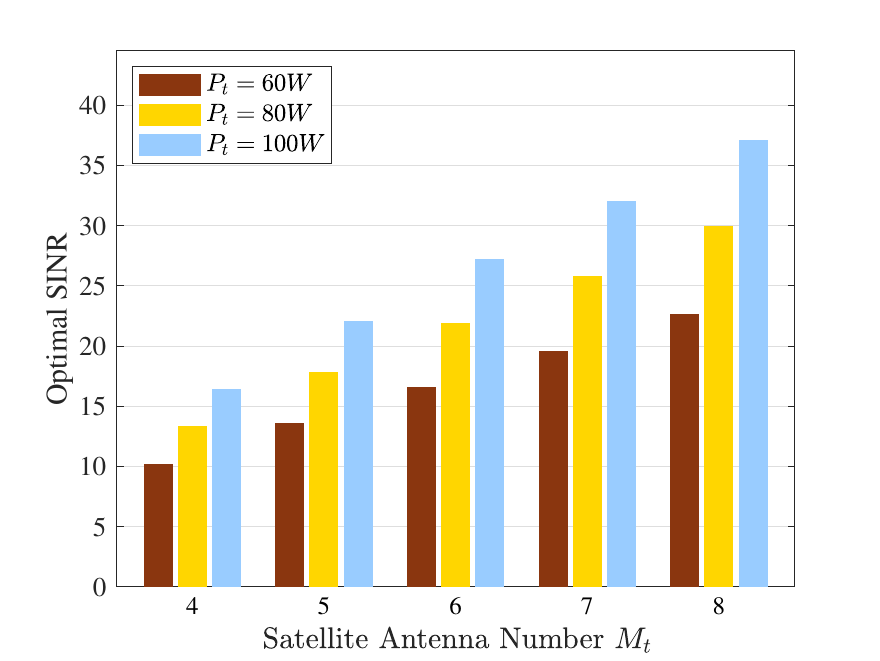}
  \vspace{-3mm}
  \renewcommand\figurename{Fig.}
  \caption{\footnotesize Optimal SINR \textit{VS} satellite antenna number.}
  \label{fig3}
\end{minipage}
\vspace{-1mm}
\end{figure*}

Fig. \ref{fig1} presents the SINR performance of the received echo signal at the filter output. It is obvious that the SINR increases automatically and converges after only 3 iterations, which validates the efficiency of the proposed algorithm. Additionally, comparing the proposed algorithm with benchmarks, we notice that by optimizing the receive filtering vector alone, a relatively high SINR value can be reached, but the maximum value cannot be obtained. However, optimizing the transmit beamforming vector alone, though it works, has incremental contributions toward the overall performance.
This is because the effect of optimization over the transmit beamforming vector can be weakened by the large-scale fading, while the ground receiver's local optimization over the receive filtering vector is dominant.

In Fig. \ref{fig2}, the optimal SINR under the bistatic mode is compared with that under the transceiving mode at different satellite altitudes, and that by optimizing the receive filtering vector alone. In all cases, the optimal SINR declines as the satellite altitude increases. 
However, it is noted that the optimization under the bistatic mode can bring a significantly higher SINR than that under the transceiving mode. This has clearly emphasized the effectiveness of using the bistatic mode rather than the transceiving mode for sensing in SAGIN.
Moreover, the SINR by optimizing the receive filtering vector alone is lower than that by the joint optimization algorithm, while the gap in between narrows as the satellite altitude increases. 
This can be due to the fact that the path loss over a longer distance gradually weakens the effect of the transmit beamforming optimization.

Fig. \ref{fig3} describes the optimal SINR against different numbers of satellite antennas under different transmit power. It is clear that with a fixed antenna number, the increase of the transmit power will brings higher SINR. 
Moreover, for different transmit power, the growth rate of SINR when increasing the antenna number can be different; the more power there is, the higher growth rate of SINR will be.
This is because adding more beams may additionally mitigate interference power, and the rise in desired signal due to increased transmit power is not directly proportional to the increase in interference power.
This provides valuable network design insights for the balance of the satellite antenna number and the transmit power when resources are limited.

% \vspace{-1mm}
\section{Conclusion}
In this paper, we studied for the first time the sensing signal performance in the SAGIN, where the LEO satellite and the terrestrial BS were separately used as a transmitter and a radar receiver to construct a bistatic sensing system. Following this setup, we jointly optimized the transmit beamforming vector and the receive filtering vector to maximize the SINR of echo signal from the target, while the closed-form expressions of both vectors were returned. Simulation results validated that the proposed algorithm can effectively improve the sensing performance in the SAGIN; meanwhile, it is manifested that the overall performance can be significantly influenced by the satellite altitude, transmit power and antenna numbers.

% \vspace{-1mm}
\section{Acknowledgement}
This work was supported in part by the YongRiver Scientific and Technological Innovation Project under Grant 2023A-187-G.
The work of Qingqing Wu was supported by the National Natural Science Foundation of China under Grant 62371289 and Grant 62331022. 

% \vspace{-1mm}
\bibliographystyle{IEEEtran}
\bibliography{references.bib}

\begin{thebibliography}{10}
\providecommand{\url}[1]{#1}
\csname url@samestyle\endcsname
\providecommand{\newblock}{\relax}
\providecommand{\bibinfo}[2]{#2}
\providecommand{\BIBentrySTDinterwordspacing}{\spaceskip=0pt\relax}
\providecommand{\BIBentryALTinterwordstretchfactor}{4}
\providecommand{\BIBentryALTinterwordspacing}{\spaceskip=\fontdimen2\font plus
\BIBentryALTinterwordstretchfactor\fontdimen3\font minus \fontdimen4\font\relax}
\providecommand{\BIBforeignlanguage}[2]{{%
\expandafter\ifx\csname l@#1\endcsname\relax
\typeout{** WARNING: IEEEtran.bst: No hyphenation pattern has been}%
\typeout{** loaded for the language `#1'. Using the pattern for}%
\typeout{** the default language instead.}%
\else
\language=\csname l@#1\endcsname
\fi
#2}}
\providecommand{\BIBdecl}{\relax}
\BIBdecl

\bibitem{yazar20206g}
A.~Yazar, S.~Dogan-Tusha, and H.~Arslan, ``6g vision: An ultra-flexible perspective,'' \emph{ITU J. Future Evol. Technol.}, vol.~1, no.~1, pp. 121--140, Jan. 2020.

\bibitem{li2024sensing}
X.~Li, S.~Guo, and S.~Malik, ``Sensing-aided peer-to-peer millimeter-wave communication,'' \emph{Int. J. Inf. Technol.}, vol.~16, no.~4, pp. 2069--2075, Apr. 2024.

\bibitem{shrestha2021survey}
R.~Shrestha, I.~Oh, and S.~Kim, ``A survey on operation concept, advancements, and challenging issues of urban air traffic management,'' \emph{Frontiers Future Transp.}, vol.~2, p.~1, 2021.

\bibitem{meng2023intelligent}
K.~Meng, Q.~Wu, C.~Masouros, W.~Chen, and D.~Li, ``Intelligent surface empowered integrated sensing and communication: From coexistence to reciprocity,'' \emph{arXiv:2311.00418}, 2023.

\bibitem{10530195}
B.~Shang, X.~Li, Z.~Li, J.~Ma, X.~Chu, and P.~Fan, ``Multi-connectivity between terrestrial and non-terrestrial mimo systems,'' \emph{IEEE Open J. Commun. Soc.}, vol.~5, pp. 3245--3262, 2024.

\bibitem{shang2023coverage}
B.~Shang, X.~Li, C.~Li, and Z.~Li, ``Coverage in cooperative leo satellite networks,'' \emph{J. Commun. Inf. Netw.}, vol.~8, no.~4, pp. 329--340, Dec. 2023.

\bibitem{liu2024max}
Z.~Liu, L.~Yin, W.~Shin, and B.~Clerckx, ``Max-min fair energy-efficient beam design for quantized isac leo satellite systems: A rate-splitting approach,'' \emph{arXiv:2402.09253}, 2024.

\bibitem{cheng2022integrated}
X.~Cheng, D.~Duan, S.~Gao, and L.~Yang, ``Integrated sensing and communications (isac) for vehicular communication networks (vcn),'' \emph{IEEE Internet Things J.}, vol.~9, no.~23, pp. 23\,441--23\,451, Dec. 2022.

\bibitem{zhang2023smart}
J.~Zhang, Q.~Yan, X.~Zhu, and K.~Yu, ``Smart industrial iot empowered crowd sensing for safety monitoring in coal mine,'' \emph{Digit. Commun. Netw.}, vol.~9, no.~2, pp. 296--305, Apr. 2023.

\bibitem{lary2016machine}
D.~J. Lary, A.~H. Alavi, A.~H. Gandomi, and A.~L. Walker, ``Machine learning in geosciences and remote sensing,'' \emph{Geosci. Front.}, vol.~7, no.~1, pp. 3--10, 2016.

\bibitem{messer2006environmental}
H.~Messer, A.~Zinevich, and P.~Alpert, ``Environmental monitoring by wireless communication networks,'' \emph{Science}, vol. 312, no. 5774, pp. 713--713, 2006.

\bibitem{ma2019wifi}
Y.~Ma, G.~Zhou, and S.~Wang, ``Wifi sensing with channel state information: A survey,'' \emph{ACM Comput. Surv.}, vol.~52, no.~3, pp. 1--36, Jun. 2019.

\bibitem{zhang2023intelligent}
Z.~Zhang, W.~Chen, Q.~Wu, Z.~Li, X.~Zhu, and J.~Yuan, ``Intelligent omni surfaces assisted integrated multi-target sensing and multi-user mimo communications,'' \emph{IEEE Trans. Commun.}, pp. 1--1, 2024.

\bibitem{tian2023active}
T.~Tian, H.~Deng, G.~Li, and R.~Ma, ``Active-passive beamforming with imperfect csi for irs-assisted sensing system,'' \emph{IEEE Signal Process. Lett.}, vol.~30, pp. 1052--1056, Aug. 2023.

\bibitem{you2022beam}
L.~You, X.~Qiang, C.~G. Tsinos, F.~Liu, W.~Wang, X.~Gao, and B.~Ottersten, ``Beam squint-aware integrated sensing and communications for hybrid massive mimo leo satellite systems,'' \emph{IEEE J. Sel. Areas Commun.}, vol.~40, no.~10, pp. 2994--3009, Oct. 2022.

\bibitem{huang2024secure}
M.~Huang, F.~Gong, G.~Li, N.~Zhang, and Q.-V. Pham, ``Secure precoding for satellite noma-aided integrated sensing and communication,'' \emph{IEEE Internet Things J.}, pp. 1--1, 2024.

\bibitem{liu2020joint}
F.~Liu, C.~Masouros, A.~P. Petropulu, H.~Griffiths, and L.~Hanzo, ``Joint radar and communication design: Applications, state-of-the-art, and the road ahead,'' \emph{IEEE Trans. Commun.}, vol.~68, no.~6, pp. 3834--3862, Jun. 2020.

\bibitem{guo2020weighted}
H.~Guo, Y.-C. Liang, J.~Chen, and E.~G. Larsson, ``Weighted sum-rate maximization for reconfigurable intelligent surface aided wireless networks,'' \emph{IEEE Trans. Wireless Commun.}, vol.~19, no.~5, pp. 3064--3076, May 2020.

\bibitem{dampahalage2021weighted}
D.~L. Dampahalage, K.~S. Manosha, N.~Rajatheva, and M.~Latva-Aho, ``Weighted-sum-rate maximization for an reconfigurable intelligent surface aided vehicular network,'' \emph{IEEE Open J. Commun. Soc.}, vol.~2, pp. 687--703, 2021.

\bibitem{shen2018fractional}
K.~Shen and W.~Yu, ``Fractional programming for communication systems—part ii: Uplink scheduling via matching,'' \emph{IEEE Trans. Signal Process.}, vol.~66, no.~10, pp. 2631--2644, May 2018.

\bibitem{boyd2004convex}
S.~P. Boyd and L.~Vandenberghe, \emph{Convex optimization}.\hskip 1em plus 0.5em minus 0.4em\relax Cambridge, U.K.: Cambridge Univ. Press, 2004.

\end{thebibliography}

\end{document}